\documentstyle[11pt,psfig]{article}    
\begin{document}
\renewcommand{\thepage}{ }
\begin{titlepage}
\title{
{\normalsize \hfill LPENSL-TH 05/2000}
{\center \bf 
Spin accumulation\\
in the semi classical and quantum regimes}}
\author{     
R. M\'elin$^{(1,2)}$ and D. Denaro$^{(2)}$ \\
{}\\
{$^{(1)}$ Centre de Recherches sur les Tr\`es basses
temp\'eratures (CRTBT)\thanks{U.P.R. 5001 du CNRS,
Laboratoire conventionn\'e avec l'Universit\'e Joseph Fourier
}}\\
{CNRS, BP 166X, 38042 Grenoble Cedex, France}\\
{}\\     
{$^{(2)}$ Laboratoire de Physique\thanks{U.M.R. CNRS 5672},
Ecole Normale Sup\'erieure de Lyon}\\
{46 All\'ee d'Italie, 69364 Lyon Cedex 07, France}\\
{}\\  
}
\date{}
\maketitle
\begin{abstract}
\normalsize   
We consider spin accumulation at a   
ferromagnet--normal metal interface in the presence of 
magnetic scattering in the normal metal.
In the classical regime, we discuss the inverse
Drude scaling of the conductance as a function of the
interface transparencies.
We present a treatment based on an exact solution of
the Boltzmann equation.
In the quantum regime, we
solve a single impurity ``spin-flip Fabry Perot
interferometer'' for quantum coherent
multiple scatterings, in which we find a resonance
in the spin flip channels. This resonance
appears to be the quantum analog of the semi classical
inverse Drude scaling of the conductance.
\end{abstract}
\end{titlepage}
\newpage
\renewcommand{\thepage}{\arabic{page}}
\setcounter{page}{1}
\baselineskip=17pt plus 0.2pt minus 0.1pt

\newpage                                       

\section{Introduction}
The discovery of the Giant Magneto Resistance (GMR)
in magnetic multilayers~\cite{exp1,exp2,exp3}
has generated an important interest.
These systems are made of a sandwich of alternating
ferromagnetic and non magnetic layers. Valet and
Fert proposed a semi classical description of the
perpendicular GMR, on the basis of a Boltzmann
equation incorporating a spin-dependent transport
in the presence of spin accumulation~\cite{Valet}
(see also~\cite{Bauer}).
Spin accumulation occurs in the GMR because
the current arising from a ferromagnet is spin polarized,
and therefore cannot penetrate a ferromagnet with an opposite
magnetization. Instead, spin accumulates at the interface.
This phenomenon occurs also at the interface between
a ferromagnet and a superconductor, where a spin
polarized current cannot penetrate the
superconductor~\cite{Falko,Jedema,Belzig}.
Here, we would like to reconsider two particular aspects
of spin accumulation, namely, (i) in the semi classical
regime, the possibility of
an inverse Drude scaling of the conductance
meaning that, in some parameter range, the conductance
increases with the length of the conductor;
and (ii) in the quantum coherent regime, the existence of
a resonance in the spin flip channels.
More precisely, we study 
a ferromagnet - normal metal - ferromagnet
spin valve, in which we assume the presence of
magnetic scattering in the normal metal~\cite{Gu98,Chen98}.
The inverse Drude scaling resulting from spin accumulation
is already implicitly contained
in the equations obtained by Valet and Fert~\cite{Valet},
but, to our knowledge, this effect has not been studied
previously {\sl per se} in the literature, which we do here.
Spin accumulation corresponds to the presence
of a different chemical potential for the spin-up
and spin-down electrons, which obey
a spin diffusion
equation~\cite{Valet,Johnson,vanSon}.
Our treatment is not based on the spin diffusion equation,
but relies on an exact solution of the 1D Boltzmann
equation where we can make an exact decoupling between the charge and
spin sectors.

Next, we ask to what extend a quantum model can
show a similar physics. We are lead to study a
quantum ``spin-flip Fabry Perot'' interferometer in which
a single magnetic impurity is located at a given
distance $a$ away from a ferromagnet interface.
We find the existence of a Fabry-Perot
resonance in the spin-flip channels as
the parameter $a$ is varied.
This resonance disappears as the ferromagnet
spin polarization is decreased, and can therefore
be viewed as the equivalent of the inverse Drude behavior
in the quantum coherent regime. Our treatment is based
on a Landauer approach, similar to the one used
by Zhu and Wang to study the effect of magnetic
scattering close to a superconductor interface~\cite{Zhu97}.

The article is organized as follows. Section~\ref{sec:Boltzmann}
is devoted to the solution of the semi classical transport
equations.
We solve the ``spin-flip Fabry Perot'' interferometer
model in section~\ref{sec:Landauer}.
Final remarks are given in the Conclusion.

\section{Transport in the semi classical regime}
\label{sec:Boltzmann}
\subsection{Boltzmann equation and boundary conditions}

\begin{figure}
\centerline{\psfig{file=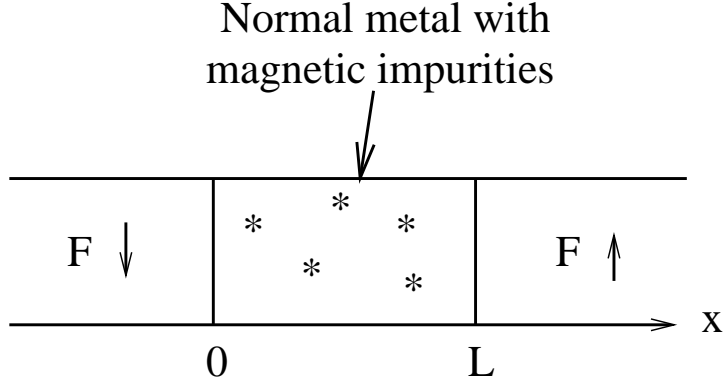,height=5cm}}
\caption{The ``spin valve'' geometry
considered in section~\ref{sec:Boltzmann},
consisting of a normal metal wire doped with magnetic
impurities, connected to two ferromagnets with
an opposite magnetization.
}
\label{fig:schema}
\end{figure}

We consider a model in which magnetic impurities
are present in a normal metal close to a
ferromagnet interface (see Fig.~\ref{fig:schema}).
We neglect
any Kondo correlation~\cite{Hewson93}, which is an
assumption valid
above the Kondo temperature.
The presence of the ferromagnets close to the normal
metal may lead to magnetic flux
lines penetrating inside the normal metal, which can
orient the magnetic impurities in a preferential direction.
We implicitly assume that the temperature is high enough
so that the impurities have no preferential orientation.
We consider a one dimensional model because only
in this geometry can we decouple the spin and charge sectors
of the Boltzmann equation. 
We assume that the interfaces between the 
ferromagnets and the normal metal are sharp,
and that the exchange
field has a step function variation at the interface. 
We note
$f_{R,L}^{\sigma}(E,x)$ the
semi classical distribution function
of right/left moving
spin-$\sigma$ electrons with an energy $E$ at position $x$.

The Boltzmann equation in the relaxation time approximation
reads
\begin{equation}
\label{eq:Boltzmann}
\frac{\partial}{ \partial x}
\left( 
\begin{array}{c}
f_R^{\uparrow} (E,x) \\
f_L^{\uparrow} (E,x) \\
f_R^{\downarrow}  (E,x)\\
f_L^{\downarrow} (E,x)
\end{array}
\right) = \left(
\begin{array}{cccc}
-(r + r_s + r_s') & r & r_s & r_s' \\
-r & r + r_s+r_s' & -r_s' & - r_s \\
r_s & r_s' & - (r + r_s + r_s') & r \\
- r_s' & -r_s & -r & r + r_s + r_s'
\end{array} \right)
\left( \begin{array}{c}
f_R^{\uparrow}  (E,x)\\
f_L^{\uparrow} (E,x) \\
f_R^{\downarrow}  (E,x)\\
f_L^{\downarrow} (E,x)
\end{array}
\right)
,
\end{equation}
where we have discarded the term involving the
electric field. This  is valid if
the temperature of the electrodes is larger than
the applied voltage, in which case the electronic gas
has a temperature identical to the one
of the electrodes~\cite{Kozub95,Nagaev95}. Therefore,
we should consider a finite temperature and calculate
the low voltage conductance in the regime $eV \ll T$.
In practise, we consider the limit $T \rightarrow 0$,
and calculate the linear conductance. 
The coefficients $r$, $r_s$ and $r_s'$ in Eq.~\ref{eq:Boltzmann}
denote respectively 
the rate of backscattering without spin-flip,
the rate of forward scattering with spin-flip,
and the rate of backward scattering with spin-flip.
The coefficients can be related to the $q=0$
and $q=2 k_f$ components of the microscopic
scattering potential (see the Appendix).

We now explicit the boundary conditions.
For this purpose, let us consider an interface
between a ferromagnet in the region $x<0$ and
a normal metal in the region $x>0$, and
include interface scattering
under the form of repulsive potential
$H \delta(x)$~\cite{Blonder82}. Let us first consider
a spin-up electron incoming from the left
ferromagnet, and denote by $b^{\uparrow}$ and $t^{\uparrow}$
the backscattering and transmission coefficients.
The wave function in the region $x<0$ is
$\psi_L(x) = \exp{(i k^{\uparrow} x)}
+ b^{\uparrow} \exp{(-i k^{\uparrow} x)}$,
and the wave function
in the region $x>0$ is $\psi_R(x) = t^{\uparrow}
\exp{(i k x)}$. The matching equations are
$\psi_L(0) = \psi_R(0) = \psi(0)$ and
$\partial \psi_R(0) / \partial x -
\partial \psi_L(0) / \partial x = (2 m H / \hbar^2)
\psi(0)$, from what we deduce
$$
t^{\uparrow} = \frac{ 2 i k^{\uparrow}}
{ i(k + k^{\uparrow}) - 2 m H / \hbar^2}
\mbox{, and }
b^{\uparrow} = \frac{ i (k^{\uparrow} - k) + 2 m H / \hbar^2}
{ i(k + k^{\uparrow}) - 2 m H / \hbar^2}
.
$$
The probability current conservation can be
verified easily: $k^{\uparrow} = k^{\uparrow}
|b^{\uparrow}|^2 + k |t^{\uparrow}|^2$.
The spin-up conductance is found to be
$G^{\uparrow} = (e^2/h) |T^{\uparrow}|^2$, with
the transmission coefficient
\begin{equation}
\label{eq:Tup}
T^{\uparrow} = \frac{k}{k^{\uparrow}} |t^{\uparrow}|^2
= \frac{4 k k^{\uparrow}}
{ (k+k^{\uparrow})^2 + \left[ \frac{2 m H}
{\hbar^2} \right]^2}
.
\end{equation}
The backscattering coefficient is $B^{\uparrow}
= 1 - T^{\uparrow}$. In the spin-down sector,
we obtain $T^{\downarrow}$ and $B^{\downarrow}$
by substituting $k^{\uparrow}$ with
$k^{\downarrow}$ in Eq.~\ref{eq:Tup}.
This provides the boundary conditions for the
Boltzmann equation:
\begin{eqnarray}
\label{eq:bound1}
f_R^\uparrow(E,0) &=& T^\downarrow f_T(E-eV)
+ (1 - T^\downarrow) f_L^\uparrow(E,0) \\
\label{eq:bound2}
f_R^\downarrow(E,0) &=& T^\uparrow f_T(E-eV)
+ (1 - T^\uparrow) f_L^\downarrow(E,0) \\
\label{eq:bound3}
f_L^\uparrow(E,L) &=& T^\uparrow f_T(E)
+ (1 - T^\uparrow) f_R^\uparrow(E,L) \\
\label{eq:bound4}
f_L^\downarrow(E,L) &=& T^\downarrow f_T(E)
+ (1 - T^\downarrow) f_R^\downarrow(E,L)
,
\end{eqnarray}
where the left and right ferromagnets are
assumed to be in equilibrium and $f_T(E)$
denotes the Fermi-Dirac distribution function.
We consider $T^\uparrow$ and $T^\downarrow$
to be independent of energy, which amounts to
considering the wave vectors $k$ and $k^\uparrow$
in Eq.~\ref{eq:Tup} to be on the Fermi surface.
Eqs.~\ref{eq:bound1}~--~\ref{eq:bound4}
provide a simple form for the spin-up and spin-down currents
at positions $x=0,L$. For instance at $x=L$, we have
\begin{eqnarray}
\label{eq:current1}
I^\uparrow(L) &=& T^\uparrow \frac{e}{h}\int  \left[ 
f_R^\uparrow(E,L) - f_T(E) \right] dE\\
I^\downarrow(L) &=& T^\downarrow \frac{e}{h} \int 
\left[ f_R^\downarrow(E,L) - f_T(E) \right]
\label{eq:current2}
dE
.
\end{eqnarray}
The Boltzmann equation Eq.~\ref{eq:Boltzmann}
and the boundary conditions
Eqs.~\ref{eq:bound1}--~\ref{eq:bound4}
lead to eight equations for eight variables
$f_{R,L}^{\uparrow,\downarrow}(E,x=0,L)$.
We now solve these equations directly and
discuss their physics.

\subsection{Solution of the Boltzmann equation}

The 4~$\times$~4 Boltzmann equation
can block diagonalized into 2~$\times$~2 blocks
by changing variables to
the charge and spin combinations
$X_{R,L} = f_{R,L}^{\uparrow} + f_{R,L}^{\downarrow}$
and $Y_{R,L} = f_{R,L}^{\uparrow} - f_{R,L}^{\downarrow}$.
This spin-charge decoupling
allows to solve exactly the Boltzmann equation.
We find
\begin{equation}
\label{eq:block}
\frac{ \partial}{ \partial x} \left(
\begin{array}{c}
X_R \\
X_L
\end{array} \right)
= \frac{1}{l} \left(
\begin{array}{cc}
-1 & 1 \\
-1 & 1 
\end{array} \right)
\left( \begin{array}{c} X_R \\ X_L
\end{array} \right)
\mbox{ , and }
\frac{ \partial}{ \partial x} \left(
\begin{array}{c}
Y_R \\
Y_L
\end{array} \right)
=  \left(
\begin{array}{cc}
-a & b \\
-b & a 
\end{array} \right)
\left( \begin{array}{c} Y_R \\ Y_L
\end{array} \right)
,
\end{equation}
with $l=1/(r + r_s')$ the mean free path,
and $a=r + 2 r_s + r_s'$,
$b = r - r_s'$. The 2~$\times$~2 block equations can
be easily integrated to obtain
\begin{equation}
\left( \begin{array}{c} X_R(L) \\ X_L(L) \end{array}
\right)
= \left( \begin{array}{cc} 1 - x & x \\
-x & 1 + x \end{array} \right)
\left( \begin{array}{c} X_R(0) \\ X_L(0) \end{array}
\right)
,
\label{eq:X-bound}
\end{equation}
with $x=L/l$. Similarly,
\begin{equation}
\label{eq:Y-bound}
\left( \begin{array}{c} Y_R(L) \\ Y_L(L) \end{array}
\right) = \hat{T}
\left( \begin{array}{c} X_R(0) \\ X_L(0) \end{array}
\right)
\mbox{ , with }
\hat{T} = \left( \begin{array}{cc}
t & u \\ - u & \overline{t}
\end{array} \right),
\end{equation}
where $t = \cosh{(\lambda L)} -
\alpha  \sinh{(\lambda L)}$,
$\overline{t} = \cosh{(\lambda L)} + \alpha \sinh{(\lambda L)}$,
and $u = \beta \sinh{(\lambda L)}$. We used the notation
$\alpha = a / \lambda$, $\beta = b / \lambda$, and
$\lambda = \sqrt{a^2 - b^2}$.
Next, we combine the boundary conditions
Eqs.~\ref{eq:bound1}--~\ref{eq:bound4} to
Eq.~\ref{eq:X-bound} to obtain an expression
for $f_R^\uparrow(E,0) - f_R^\downarrow(E,0)$
and $f_L^\uparrow(E,0) - f_L^\downarrow(E,0)$
as a function of $f_R^\uparrow(E,L)$ and
$f_R^\downarrow(E,L)$. Once injected 
into Eq.~\ref{eq:Y-bound}, these relations 
lead to 
\begin{eqnarray}
\label{eq:solu}
\hat{M} \left( \begin{array}{c} f_R^\uparrow(E,L) \\
f_R^\downarrow(E,L) \end{array} \right) &=&
2 f_T(E-eV) \hat{T} \left( \begin{array}{c}
T^\uparrow + T^\downarrow - 2 T^\uparrow T^\downarrow \\
T^\uparrow + T^\downarrow \end{array} \right)\\
\nonumber
&-&  f_T(E) \left\{ (T^\uparrow + T^\downarrow)
\hat{T} \left( \begin{array}{c}
2 (T^\uparrow + T^\downarrow - T^\uparrow T^\downarrow -1)
-x (T^\uparrow + T^\downarrow - 2 T^\uparrow T^\downarrow) \\
-2 + T^\uparrow + T^\downarrow - x(T^\uparrow + T^\downarrow)
\end{array} \right) \right. \\
\nonumber
 &-& \left.(T^\uparrow - T^\downarrow)^2 \left( \begin{array}{c}
0 \\ 1 \end{array} \right)  \right\}
.
\end{eqnarray}
The matrix $\hat{M}$ appearing in the left hand side of
Eq.~\ref{eq:solu} is
\begin{equation}
\label{eq:Mhat}
\hat{M} = \hat{T} \left( \begin{array}{cc}
A^\uparrow & A^\downarrow \\
B^\uparrow & B^\downarrow \end{array} \right)
+ (T^\uparrow - T^\downarrow)
\left( \begin{array}{cc} -1 & 1 \\
-1+T^\uparrow & 1 - T^\downarrow \end{array} \right)
,
\end{equation}
with the coefficients
\begin{eqnarray}
\label{eq:Auparrow}
A^\uparrow &=& 3 T^\uparrow - 4 T^\uparrow T^\downarrow
+ T^\downarrow - 2 (T^\uparrow)^2 + 2 (T^\uparrow)^2
T^\downarrow
+ x T^\uparrow \left[ T^\uparrow - 2 T^\uparrow
T^\downarrow + T^\downarrow \right]\\
B^\uparrow &=& 3 T^\uparrow + T^\downarrow
- (T^\uparrow)^2 - T^\uparrow T^\downarrow
+ x \left[ (T^\uparrow)^2 + T^\uparrow T^\downarrow \right]
\label{eq:Buparrow}
.
\end{eqnarray}
The expression of $A^\downarrow$ is obtained by
exchanging $T^\uparrow$ and $T^\downarrow$
in Eq.~\ref{eq:Auparrow}.
Similarly, $B^\downarrow$
is obtained from $B^\uparrow$ 
by exchanging
$T^\uparrow$ and $T^\downarrow$
in Eq.~\ref{eq:Buparrow}.

\subsection{Fully polarized limit}
We first consider the solution Eq.~\ref{eq:solu} in the
case of fully polarized ferromagnets with high transparency
contacts: $H=0$,
$k^\downarrow = T^\downarrow =0$,
$k^\uparrow = k$, leading to $T^\uparrow=1$.
In this limit, only a spin-up current can enter
the ferromagnet at $x=L$. This is expected
on physical grounds, and it can be verified
explicitly on the form Eq.~\ref{eq:current2}
of the spin-down current.
The total current is found to be
\begin{equation}
\label{eq:I-fully}
I = \frac{e}{h} \int dE
\frac{2 \left[ f_T(E-eV) - f_T(E) \right]
\sinh{\left[ 2 \sqrt{r_s(r_s+r)} L\right]}}
{(2 + rL) \sinh{ \left[ 2 \sqrt{r_s(r_s+r)} L\right]}
+ \sqrt{ \frac{r}{r_s} + 1}
\left\{ \cosh{\left[ 2 \sqrt{r_s(r_s+r)} L\right]}
+ 1 \right\}}
,
\end{equation}
where we considered only the forward scattering
spin flip processes ($r_s'=0$), and assumed that
$r_s \ll r$, in which case
the elastic mean free path $l=1/r$
is much below the spin-flip length
$l_{\rm sf} = 1/[2 \sqrt{r_s(r_s+r)}]$.
If $L$ is small compared to $l_{\rm sf}$,
the conductance
$ G \sim 2 \frac{e^2}{h} r_s L $
shows an inverse Drude behavior.

\subsection{Spin polarization profile}
\begin{figure}
\centerline{\psfig{file=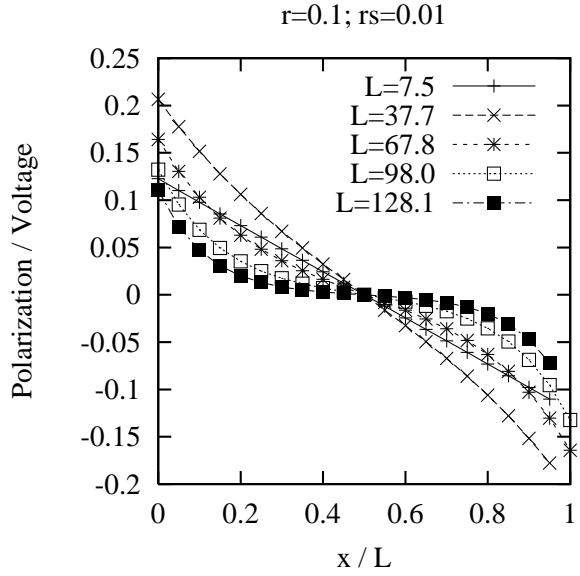,height=8cm}}
\caption{Spin polarization profile inside the diffusive
conductor, with $r=0.1$, $r_s=0.01$, and increasing
values of $L$. The spin polarization is normalized
to the applied voltage, and the coordinate along the
wire is normalized to the total length. We have
$1/\lambda = 15.8$. 
}
\label{fig:pol-wire}
\end{figure}

The spin polarization profile in the diffusive wire
can be calculated in a straightforward fashion from
the solution of the Boltzmann equation.
Once we know the distribution functions at one
extremity of the wire, we can
use Eqs.~\ref{eq:block} to
propagate the solution to an arbitrary point.
The resulting spin
polarization inside the wire is proportional to
the applied voltage, and is shown on Fig.~\ref{fig:pol-wire}
for various values of $L$. When $L > l_{\rm sf}$,
there is a plateau
in the spin polarization in the middle of the wire.
In the opposite inverse Drude regime, there is no such
plateau.

\begin{figure}
\centerline{\psfig{file=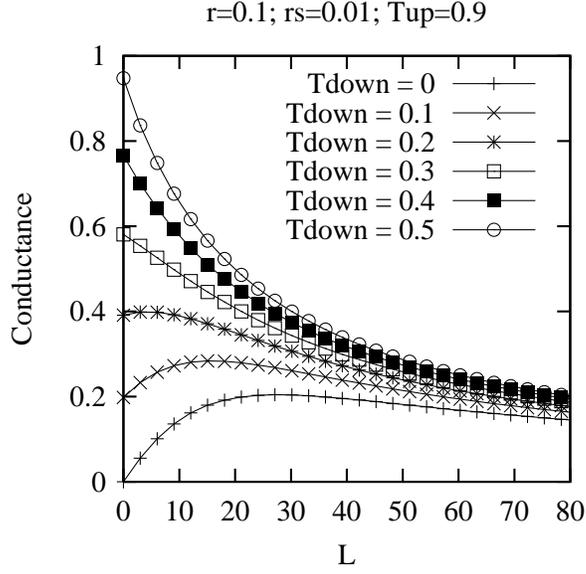,height=8cm}}
\caption{Variations of the conductance as a
function of the length $L$ of the diffusive wire,
with decreasing spin polarizations $T^\downarrow=
0$, $0.1$, $0.2$, $0.3$, $0.4$, $0.5$, and the parameters
$T^\uparrow=0.9$,
$r=0.1$ and $r_s=0.01$. With a strong spin
polarization, the conductance increases with
$L$ below $L_c$. With a weak spin polarization, the conductance
decreases monotonically with $L$ ($L_c=0$).
}
\label{fig:G-finite}
\end{figure}
\begin{figure}
\centerline{\psfig{file=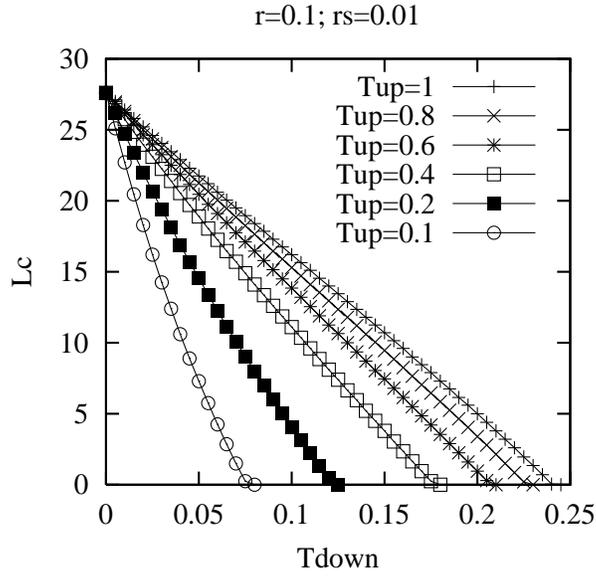,height=8cm}}
\caption{Variations of $L_c$ as a
function of $T^\downarrow$,
with increasing spin polarizations $T^\uparrow=
0.1$, $0.2$, $0.4$, $0.6$, $0.8$, $1$, and the parameters
$r=0.1$ and $r_s=0.01$. With a given $T^\uparrow$,
the curve $L_c(T^\downarrow)$ separates two
regions: (i) small-$L$, small-$T^\downarrow$:
the conductance increases with $L$;
and (ii) large-$L$, large-$T^\downarrow$:
the conductance decreases with $L$.
}
\label{fig:Lc}
\end{figure}

\subsection{Effect of a partial spin polarization}
  
Now, we consider the effect of a partial spin polarization
in the ferromagnet. 
It is expected on physical grounds
that a decreasing spin polarization tends to suppress
the inverse Drude scaling
because this regime is clearly
absent in the spin unpolarized case. This is visible
on Fig.~\ref{fig:G-finite} where we plotted the
conductance as a function of the length of the diffusive wire
for decreasing spin polarizations. With an arbitrary
polarization, there exists a critical
length scale $L_c$ such
that the conductance increases with $L$ below $L_c$,
and decreases with $L$ above $L_c$. There
exists also a critical value of $T^\downarrow$
such that $L_c=0$ if $T^\downarrow > T^\downarrow_c$.
To illustrate this, we have shown on Fig.~\ref{fig:Lc} the
variations of the critical length $L_c$ as a function of
the parameter $T^{\downarrow}$. 
When $T^\downarrow$ increases,
$L_c$ decreases: the maximum in $G(L)$ occurs
for a smaller $L_c$. When $T^\downarrow$
is above a critical value $T^\downarrow_c$,
the conductance decreases monotonically with
$L$.

We now describe the effect of a partial
spin polarization on the basis of a
small-$T^{\downarrow}$ expansion.
The strategy is to express the current to order $L$
and determine whether the conductance increases
or decreases with $L$.
We expand the current to first order in the two parameters
$K=\lambda L$ and $x=L/l$, and retain the
coefficients of this expansion to leading
order in $T^\downarrow$.
It is first instructive
to carry out the expansion with $L=0$,
and therefore $K=x=0$.  It is visible on
Eqs.~\ref{eq:current1},~\ref{eq:current2} that
a prefactor $T^\uparrow$ enters the spin-up
current, and a prefactor $T^\downarrow$
enters the spin-down current.
We should then express $f_R^\uparrow(E,L)$
to first order in $T^\downarrow$ while
$f_R^\downarrow(E,L)$ should be
expressed to order $(T^\downarrow)^0$.
The spin-up and spin-down channels
appear to play an asymmetric role. Nevertheless,
the final expression of the conductance is
identical in the spin-up and spin-down
channels.
An intermediate step in the calculation
of $f_R^\uparrow(E,L)$ is the derivation
of $\mbox{Det} \hat{M}$
to order $T^\downarrow$ (see Eq.~\ref{eq:Mhat}):
$$
\mbox{Det} \hat{M} = - 4 (T^{\uparrow})^3
\left\{ 1 - T^\downarrow \left( \frac{2 T^\uparrow -1}
{T^\uparrow} \right) \right\}
,
$$
leading to an identical current in both spin channels:
$$
I^\uparrow = I^\downarrow
= T^\downarrow \frac{e}{h} \int \left[
f_T(E-eV) - f_T(E) \right] dE
.
$$
Now we consider a diffusive wire with a finite length $L$,
and expand the current to first order
in $x$ and $K$, and to leading order in
$T^\downarrow$.
The determinant of the
matrix $\hat{M}$ in Eq.~\ref{eq:solu} is found to be
$$
\mbox{Det} \hat{M} = -4 (T^\uparrow)^3
\left\{ 1 - T^\downarrow \left(
\frac{2 T^\uparrow-1}{T^\uparrow} \right) \right\}
- 4 (T^\uparrow)^2 (\alpha - \beta) (2 - T^\uparrow)
K - 4 x (T^\uparrow)^2 T^\downarrow
.
$$
Next we expand the spin-up current to order $L$
to obtain
$$
I^\uparrow = \frac{- 4 (T^\uparrow)^3}
{ \mbox{Det} \hat{M}} \left( T^\downarrow
- (\alpha - \beta) K \right)
\simeq T^\downarrow \left[
1 + K \frac{\alpha - \beta}{T^\downarrow}
- x \frac{T^\downarrow}{T^\uparrow} \right]
.
$$
If $T^\downarrow$ is small, the current
increases with $L$ while it decreases with $L$
if $T^\downarrow$ is large.
The transition between these two behaviors is
obtained for
$T^\downarrow_c = \sqrt{ (2 r_s/r) T^\uparrow}$,
compatible with the behavior shown on Fig.~\ref{fig:Lc}.

\subsection{Replacement of one of the ferromagnets
by a normal metal}

\begin{figure}
\centerline{\psfig{file=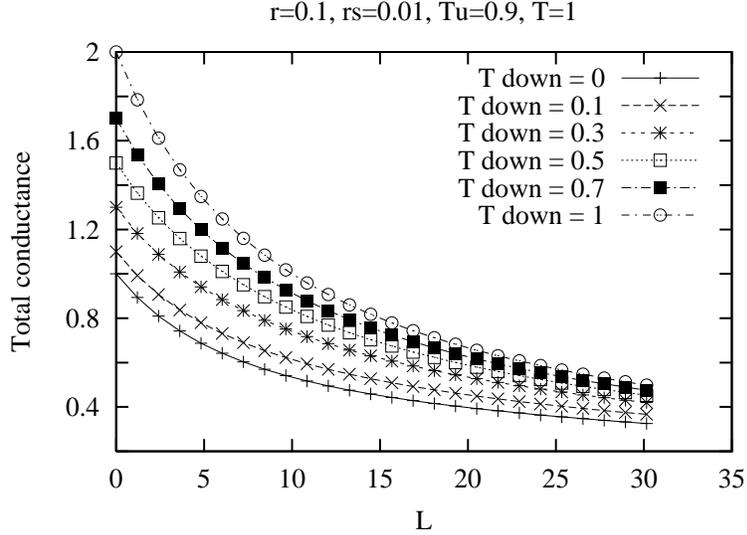,height=7.5cm}}
\caption{ Conductance of the junction with a diffusive wire
connected to a normal metal and a ferromagnet, with
high transparency contacts: $T=T^\uparrow=1$. It is
visible that the conductance decreases monotonically
with the length of the diffusive wire.
}
\label{fig:T-1}
\end{figure}

We now consider the situation where we replace the
left-hand-side ferromagnet on Fig.~\ref{fig:schema}
by a normal metal. In the presence of
high transparency contacts, the conductance of
this junction  is of order $e^2/h$ in the absence of diffusion
while it is of order $(e^2/h) T^\downarrow$
in the spin valve geometry on
Fig.~\ref{fig:schema}. Replacing one of
the ferromagnets by a normal metal is expected to 
suppress the inverse Drude scaling.
The boundary conditions appropriate to describe
this situation are
\begin{eqnarray}
f_R^\uparrow(E,0) &=& T f_T(E-e V)
+ (1-T) f_L^\uparrow(E,0) \\
f_R^\downarrow(E,0) &=& T f_T(E-e V)
+ (1-T) f_L^\downarrow(E,0) \\
f_L^\uparrow(E,L) &=& T^\uparrow  f_T(E)
+ (1-T^\uparrow) f_R^\uparrow(E,0) \\
f_L^\downarrow(E,L) &=& T^\downarrow  f_T(E)
+ (1-T^\downarrow) f_R^\downarrow(E,0) 
,
\end{eqnarray}
that should be solved together with Eqs.~\ref{eq:block}.
The solution is found to be
$$
\hat{N} \left(
\begin{array}{c} f_R^\uparrow(E,L) \\
f_R^\downarrow(E,L) \end{array} \right)
= \left( \begin{array}{c}
(T^\uparrow + T^\downarrow) (1-T +xT) \\
(T^\uparrow - T^\downarrow)(u + t(1-T))
\end{array} \right) f_T(E)
+ 2 T \left(
\begin{array}{c} 1 \\ 0 \end{array} \right)
f_T(E - e V)
,
$$
with
$$
\hat{N} = \left( \begin{array}{cc}
C^\uparrow & C^\downarrow \\
D^\uparrow & -D^\downarrow
\end{array} \right)
,
$$
and
\begin{eqnarray}
C^\uparrow &=& T + T^\uparrow - T T^\uparrow +
x T T^\uparrow \\
D^\uparrow &=& \overline{t} - u (1-T)
- (1 - T^\uparrow)( t(1-T) +u)
.
\end{eqnarray}
We have plotted on Fig.~\ref{fig:T-1} the conductance of the
junction with high transparency contacts, where it is visible
that the conductance
decreases monotonically with the length of the
diffusive wire. 

\begin{figure}
\centerline{\psfig{file=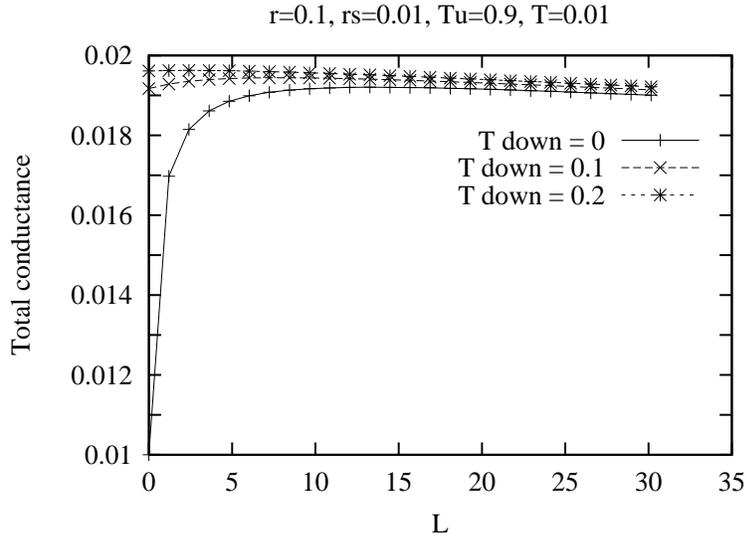,height=7.5cm}}
\caption{ Conductance of the junction with a diffusive wire
connected to a normal metal and a ferromagnet, with
low transparency contacts: $T=0.01$, and $T^\uparrow=1$. 
A regime with a conductance increasing with the length of the
diffusive wire is visible for small values of the
parameter $T^\downarrow$.
}
\label{fig:T-.01}
\end{figure}

Now, reducing the contact transparency restores a
regime in which the conductance increases with
the size of the diffusive wire. This is visible on
Fig.~\ref{fig:T-.01} where we used $T=0.01$
and $T^\uparrow=1$. Again, the inverse Drude scaling
is obtained for the smallest
values of $T^\downarrow$ (with strongly
polarized magnets).

\section{Quantum coherent transport: a single
magnetic impurity ``spin-flip Fabry Perot interferometer''}
\label{sec:Landauer}
\subsection{Matching equations}
\begin{figure}
\centerline{\psfig{file=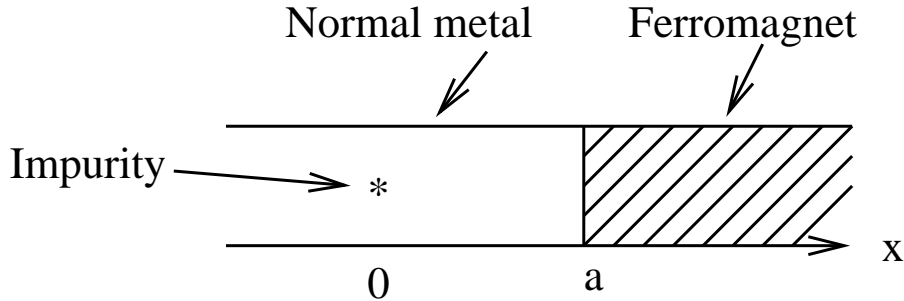,height=4cm}}
\caption{The system considered in section~\ref{sec:Landauer}.
The impurity is in the normal metal at a distance $a$
away from the ferromagnet. We have represented a quasi
one dimensional geometry while the calculation
is made in a one dimensional geometry.
}
\label{fig:schem2}
\end{figure}

We now consider a single magnetic impurity at
$x=0$ in a normal metal, in the presence of
a normal metal -- ferromagnet interface at $x=a$
(see Fig.~\ref{fig:schem2}). The purpose of this
calculation is to study a model in which the interplay between
multiple reflections and phase coherence is treated exactly,
and to determine whether there exists a signature
of spin accumulation in the quantum coherent regime.
We find that the quantum model behaves like a
Fabry Perot interferometer, with a resonance in the spin
flip channels. This can be viewed as the signature of
spin accumulation in the quantum coherent regime.
We neglect the Kondo effect because we want to describe
a situation in which the temperature is above the
Kondo temperature. The conduction electrons are scattered
through the Hamiltonian ${\cal H}=V_0
+ V_1 {\bf S}_i . {\bf s}$, where ${\bf S}_i$ is
the impurity spin and ${\bf s}$ the spin of the conduction
electron, and we further assume a single
channel geometry.
This type of model has been used by
Zhu and Wang~\cite{Zhu97} to investigate the effect of
a magnetic impurity close to a normal metal -- superconductor
interface.
The spin-up and spin-down wave functions
are grouped in a two-component spinor $\hat{\psi}(x)$.
Clearly, the impurity couples the spin-up and spin-down
wave functions.
The matching of the wave function at
the impurity site reads
\begin{equation}
\label{eq:match1}
\hat{\psi}(0^+) = \hat{\psi}(0^-)
\mbox{, and }
\frac{\partial \hat{\psi}}{\partial x} (0^+)
- \frac{\partial \hat{\psi}}{\partial x}(0^-)
= \frac{2 m}{\hbar^2} \left[ \lambda \hat{1}
+ \mu \hat{\sigma}_x \right] \hat{\psi}(0)
,
\end{equation}
with $\lambda=V_0 - V_1/4$ and $\mu=V_1/2$.
The matching of the wave function at
the ferromagnet boundary reads
\begin{equation}
\label{eq:match2}
\hat{\psi}(a^+) = \hat{\psi}(a^-)
\mbox{, and }
\frac{ \partial \hat{\psi}}{\partial x} (a^+)
- \frac{\partial \hat{\psi}}{\partial x} (a^-)
= \frac{2 m}{\hbar^2} H \hat{\psi}(a)
,
\end{equation}
where we included a repulsive interface potential
$H \delta(x-a)$ at the normal metal -- ferromagnet interface.
Eqs.~\ref{eq:match1},~\ref{eq:match2}
generate eight constraints,
for a set of eight transmission coefficients.

\begin{figure}
\centerline{\psfig{file=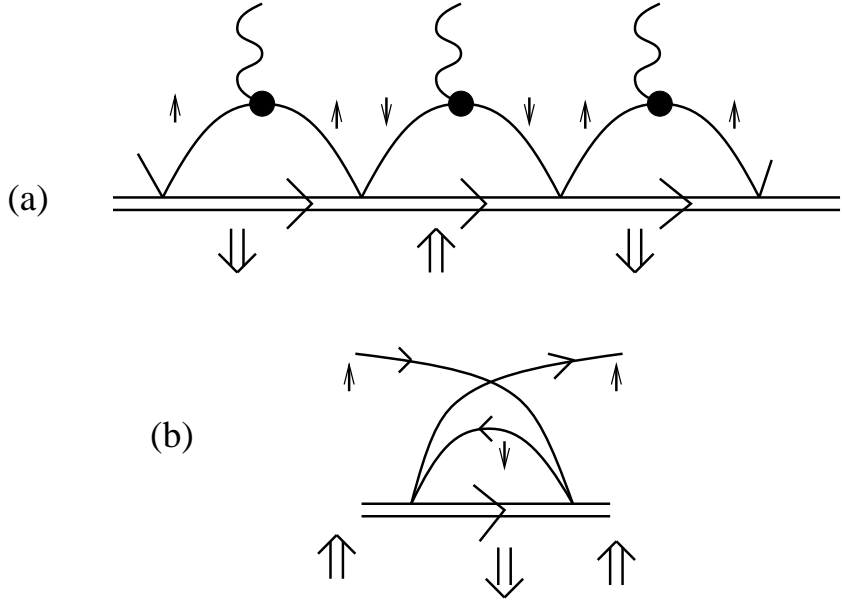,height=8cm}}
\caption{ (a) The processes included in the Landauer
calculation. The wavy lines indicate the scattering
at the interface; and (b) A process with a hole in
the intermediate state, not included in
the calculation.
}
\label{fig:diagram1}
\end{figure}

This calculation amounts to a resummation to all orders
of a series of diagrams in which a conduction electron
scatters onto the impurity, scatters back onto the
interface, scatters again onto the impurity, ...
(see Fig.~\ref{fig:diagram1}~(a)).
Note that the diagram with a hole in the intermediate
state shown on Fig.~\ref{fig:diagram1}~(b)
generates another series which is not included
in the calculation. 
If one wanted to describe the Kondo
effect close to a ferromagnet interface, it would be
crucial to incorporate the diagram on
Fig.~\ref{fig:diagram1}~(b), as well as inserting
the interface scattering in this diagram.

\subsection{Scattering in the total spin $S^z=0$ sectors}
\subsubsection{Incoming electron with a spin-up}
\label{sec:spin-up}
We first consider a spin-up electron incoming
on the interface while the impurity is supposed
to have initially a spin down.
The wave functions are
\begin{eqnarray}
\hat{\psi}^{e \uparrow}_{i \downarrow}(x) &=&
\left( \begin{array}{c} 1 \\ 0 \end{array}
\right) e^{i k x}
+ \left( \begin{array}{c}
b^{e \uparrow \rightarrow e \uparrow}_{i \downarrow}
\\ b^{e \uparrow \rightarrow e \downarrow}_{i \downarrow}
\end{array} \right)
e^{-i k x}
\mbox{ if $x<0$.}\\
\hat{\psi}^{e \uparrow}_{i \downarrow}(x) &=& \left( \begin{array}{c}
\alpha \\ \alpha' \end{array} \right) e^{i k x}
+ \left( \begin{array}{c} \beta \\ \beta'
\end{array} \right) e^{-i k x}
\mbox{ if $0 < x <a$.}\\
\hat{\psi}^{e \uparrow}_{i \downarrow}(x)
&=& t^{e \uparrow \rightarrow e \uparrow}_{i \downarrow}
\left( \begin{array}{c}1 \\ 0 \end{array}
\right) e^{i k^{\uparrow} x} +
t^{e \uparrow \rightarrow e
\downarrow}_{i \downarrow} \left(
\begin{array}{c} 0 \\ 1 \end{array} \right)
e^{i k^{\downarrow} x}
\mbox{ if $x>a$,}
\end{eqnarray}
where $k^{\uparrow}$ and $k^{\downarrow}$ denote the
spin-up and spin-down Fermi wave vectors in the ferromagnet.
In the notation of the transmission coefficients,
the superscript denotes the initial and final
spin orientations of the conduction electron while
the subscript denotes the initial orientation of
the impurity.
The solution of the matching equations is straightforward,
and we find the transmission coefficients
\begin{eqnarray}
\label{eq:solu-t}
t^{e \uparrow \rightarrow e \uparrow}_{i \downarrow}
&=& \frac{1}{ {\cal D} A^{\uparrow}}
\left[ \overline{A} (1 + i z) X^{\downarrow}
+ A i z Y^{\downarrow} \right]\\
t^{e \uparrow \rightarrow e \downarrow}_{i \downarrow}
&=& -\frac{1}{ {\cal D} A^{\downarrow}}
i z'
\left[ \overline{A} X^{\uparrow}
+ A  Y^{\uparrow} \right]
\label{eq:solu-t-bis}
.
\end{eqnarray}
We used the notation
$X^{\sigma} = {1 \over 2} + i Z + {Z \over 2
Z^\sigma}$, $Y^{\sigma} = {1 \over 2} 
- \left( i Z + {Z \over 2 Z^\sigma} \right)$,
$A = \exp{(i k a)}$, $A^\sigma = \exp{(i k^{\sigma} a)}$.
The dimensionless scattering potentials in
Eqs.~\ref{eq:solu-t},~\ref{eq:solu-t-bis}
are $z = m \lambda / (\hbar^2 k)$ and
$z' = m \mu / (\hbar^2 k)$ at the impurity
site, and $Z=m H / (\hbar^2 k)$, $Z^\sigma
= m H / (\hbar^2 k^\sigma)$ at the normal metal
-- ferromagnet interface. The denominator ${\cal D}$
in Eqs.~\ref{eq:solu-t},~\ref{eq:solu-t-bis} is
\begin{equation}
\label{eq:D}
{\cal D} = X^\uparrow X^\downarrow 
(\overline{A})^2
[ 1 - z^2 + (z')^2
+ 2 i z ]
+ (X^\uparrow Y^\downarrow + X^\downarrow Y^\uparrow)
[ - z^2 + (z')^2 + i z ]
+ Y^\uparrow Y^\downarrow A^2
[ - z^2 + (z')^2]
.
\end{equation}

\subsubsection{Incoming electron with a spin-down}
We consider now an incoming electron with a spin-down 
while the impurity has initially a spin-up.
The wave functions are
\begin{eqnarray}
\hat{\psi}^{e \downarrow}_{i \uparrow}(x) &=&
\left( \begin{array}{c} 0 \\ 1 \end{array}
\right) e^{i k x}
+ \left( \begin{array}{c}
b^{e \downarrow \rightarrow e \uparrow}_{i \uparrow}
 \\ b^{e \downarrow \rightarrow e \downarrow}_{i \uparrow}
\end{array} \right)
e^{-i k x}
\mbox{ if $x<0$.}\\
\hat{\psi}^{e \downarrow}_{i \uparrow}(x) &=& \left( \begin{array}{c}
\alpha' \\ \alpha \end{array} \right) e^{i k x}
+ \left( \begin{array}{c} \beta' \\ \beta
\end{array} \right) e^{-i k x}
\mbox{ if $0 < x <a$.}\\
\hat{\psi}^{e \downarrow}_{i \uparrow}(x) &=&
t^{e \downarrow \rightarrow e \downarrow}_{i \uparrow}
\left( \begin{array}{c}0 \\ 1 \end{array}
\right) e^{i k^{\downarrow} x} +
t^{e \downarrow \rightarrow e \uparrow}_{i \uparrow}
\left(
\begin{array}{c} 1 \\ 0 \end{array} \right)
e^{i k^{\uparrow} x}
\mbox{ if $x>a$.}
\end{eqnarray}
The equations
for $t^{e \downarrow \rightarrow e \uparrow}_{i \uparrow}$
and $t^{e \downarrow \rightarrow e \downarrow}_{i \uparrow}$
are obtained from the ones
in section~\ref{sec:spin-up} under the transformation
$ A^\uparrow \leftrightarrow A^\downarrow$, and
$Z^\uparrow \leftrightarrow Z^\downarrow$.
The amplitude for transmission in the
ferromagnet is
\begin{eqnarray}
t^{e \downarrow \rightarrow e \uparrow}_{i \uparrow}
&=& - \frac{1}{ {\cal D} A^\uparrow}
i z' [ \overline{A} X^\downarrow + A Y^\downarrow]\\
t^{e \downarrow \rightarrow e \downarrow}_{i \uparrow}
&=&  \frac{1}{ {\cal D} A^\downarrow}
[ \overline{A} (1+iz) X^\uparrow + A iz Y^\uparrow]
.
\end{eqnarray}

\subsection{Scattering in the total spin $S^z=\pm 1$ sectors}
The
incoming electron does not undergo spin-flip scattering
in the sectors with a total spin $S^z = \pm 1$.
The transmission coefficients in the sector
$S^z=1$ is found to be
$$
t^{e \uparrow \rightarrow e \uparrow}_{i \uparrow} =
\frac{1}{A^\uparrow [ \overline{A}
\left( 1 + i (z+z') \right) X^\uparrow
+ A i (z+z') Y^\uparrow ]}
.
$$
In the sector $S^z=-1$, we have
$$
t^{e \downarrow \rightarrow e \downarrow}_{i \downarrow} =
\frac{1}{A^\downarrow [ \overline{A}
\left( 1 + i (z+z') \right) X^\downarrow
+ A i (z+z') Y^\downarrow ]}
.
$$
We can check easily that these forms of the
transmission coefficients are identical to
Eq.~\ref{eq:solu-t}, with $z'=0$, and the
replacement $z \rightarrow z + z'$. This is expected
since there is no spin-dependent scattering
in the limit $z'=0$ of Eq.~\ref{eq:solu-t}.

\subsection{Landauer formula}
We now evaluate the total conductance and assume
that the incoming electron and impurity do not have any
preferential direction. The conductance is
the sum of four terms, weighted by the probability
${\cal P}=1/2$ to have a spin-up or spin-down impurity:
$G = \left(
G^{e \uparrow}_{i \downarrow} +
G^{e \downarrow}_{i \uparrow} +
G^{e \uparrow}_{i \uparrow} +
G^{e \downarrow}_{i \downarrow} \right)/2$, with
\begin{eqnarray}
G^{e \uparrow}_{i \downarrow} &=& \frac{e^2}{h}
\left(
\frac{ k^\uparrow}{k}
|t^{e \uparrow \rightarrow e \uparrow}_{i \downarrow}|^2
+ \frac{ \mbox{Re} k^\downarrow}{k}
|t^{e \uparrow \rightarrow e \downarrow}_{i \downarrow}|^2
\right) \\
G^{e \downarrow}_{i \uparrow} &=& \frac{e^2}{h}
\left(
\frac{ k^\uparrow}{k}
|t^{e \downarrow \rightarrow e \uparrow}_{i \uparrow}|^2
+ \frac{ \mbox{Re} k^\downarrow}{k}
|t^{e \downarrow \rightarrow e \downarrow}_{i \uparrow}|^2
\right) \\
G^{e \uparrow}_{i \uparrow} &=& \frac{e^2}{h}
\frac{ k^\uparrow}{k}
|t^{e \uparrow \rightarrow e \uparrow}_{i \uparrow}|^2 \\
G^{e \downarrow}_{i \downarrow} &=& \frac{e^2}{h}
\frac{ \mbox{Re}k^\downarrow}{k}
|t^{e \downarrow \rightarrow e \downarrow}_{i \downarrow}|^2 
.
\end{eqnarray}
We have incorporated the possibility of having a pure
imaginary wave vector $k^\downarrow$, corresponding
to an empty spin-down band.

\subsection{Resonances}
\begin{figure}
\centerline{\psfig{file=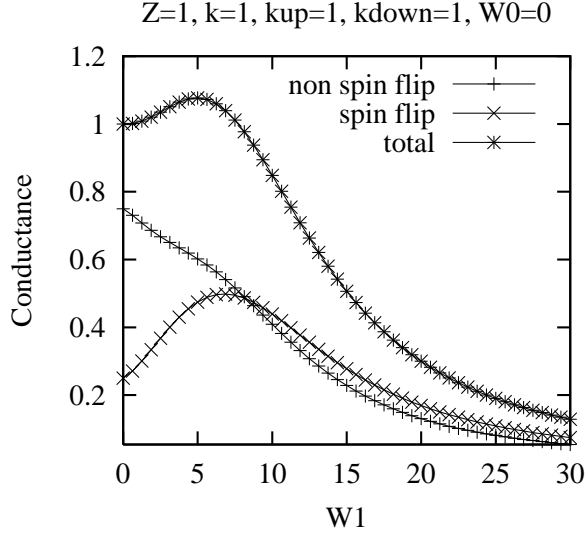,height=7.5cm}}
\caption{ Spin-flip, non spin-flip, and total conductances (in units of
$e^2/h$) of the
quantum mechanical model of magnetic scattering close to a ferromagnet
interface, with an unpolarized ferromagnet 
($k=k^\uparrow = k^\downarrow =1$) and the parameters
$Z=1$, $a=100$ and $W_0 = V_0/H=0$. The conductances
are plotted as a function of $W_1 = V_1/H$.
}
\label{fig:non-pol}
\end{figure}
\begin{figure}
\centerline{\psfig{file=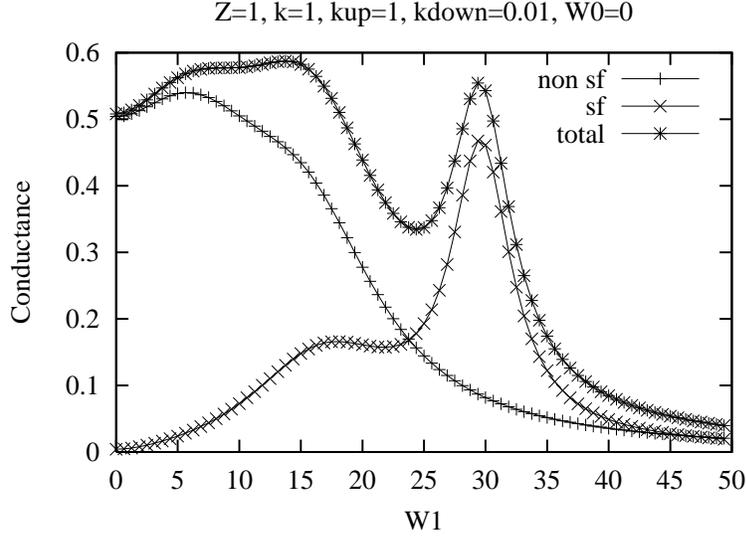,height=7.5cm}}
\caption{ Spin-flip, non spin-flip, and total conductances (in units of
$e^2/h$) of the
quantum mechanical model of magnetic scattering close to a ferromagnet
interface, with a strongly polarized ferromagnet 
($k=k^\uparrow = 1$, $k^\downarrow =0.01$) and the parameters
$Z=1$, $a=100$ and $W_0 = V_0/H=0$. The conductances
are plotted as a function of $W_1 = V_1/H$. A resonance,
not present on Fig.~\ref{fig:non-pol},  develops in the conductance
upon spin polarizing the ferromagnet.
}
\label{fig:pol}
\end{figure}

\begin{figure}
\centerline{\psfig{file=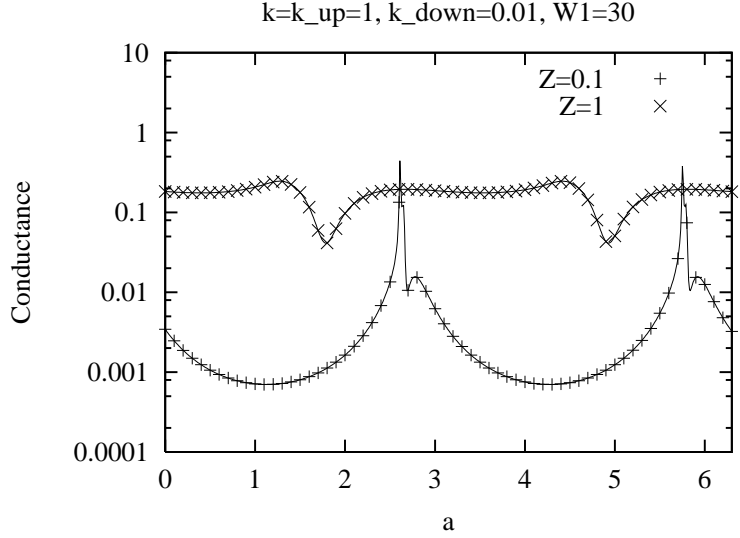,height=7.5cm}}
\caption{ Spin flip conductance of the Fabry Perot
model, as a function of $a$, with
$k=k^\uparrow=1$, $k^\downarrow=0.01$.
As it is visible, the conductance is a periodic
function of $k a$ with a period $\pi$. The resonances
with $Z=1$ are much sharper than with $Z=0.01$,
as expected on physical grounds. The resonances
occur when $\tan{(k a)} \simeq -1/(2Z)$.
}
\label{fig:shape-reso}
\end{figure}

We consider the presence
of a strong interface scattering at the metal -- ferromagnet interface.
The electron are multiply reflected
before they enter the ferromagnet, and therefore
the resonator has a high quality factor.
We first choose
the parameter $a$ in such a way that spin-flip scattering
is resonant.
As it is visible on Figs.~\ref{fig:non-pol}
and~\ref{fig:pol}, the presence of a spin polarization
in the ferromagnet
generates a resonance in the conductance, not present
in the unpolarized
situation. We have shown the conductance with $V_0=0$
but a similar behavior has been obtained with
a finite $V_0$.
The values of $ka$ for which a resonance
occurs can be worked out by calculating the transmission
coefficients in the limit of a large $z$, $z'$.
In this limit, we find
\begin{eqnarray}
\label{eq:t-large-z1}
t^{e \downarrow \rightarrow e \uparrow}_{i \uparrow}
&=& - \frac{i z'\overline{A}^\uparrow}{((z')^2 - z^2)}
\frac{1}{ \overline{A} X^\uparrow + A Y^\uparrow}\\
t^{e \uparrow \rightarrow e \uparrow}_{i \downarrow}
&=& \frac{i z \overline{A}^\uparrow}{ (z')^2 - z^2} \frac{1}{\overline{A}
X^\uparrow + A Y^\uparrow}
\label{eq:t-large-z2}
.
\end{eqnarray}
The resonances occur when $\tan{(ka)} = 1/(i-2Z)$.
For a large $Z$, the resonances are close to the real
axis: $\tan{(k a)} = -1/(2Z)$,
in agreement with Fig.~\ref{fig:shape-reso}.
The reason why the resonance appear to be sharp
as a function of $a$
when $Z$ is large is that, even without spin flip
scattering, the quality factor of such a resonator
is large when $Z$ is large.

The occurrence of a parameter range
in which a peak occurs in the spin flip conductance
is intriguing.
The
presence of a specific physics in the spin flip channels
can be already understood from the large-$z$, $z'$
behavior, Eqs.~\ref{eq:t-large-z1},~\ref{eq:t-large-z2}.
Typically, one has
$
|t^{e \downarrow \rightarrow e \uparrow}_{i \uparrow}|^2
\sim (z')^2 / ((z')^2-z^2)^2$ and
$|t^{e \uparrow \rightarrow e \uparrow}_{i \uparrow}|^2
\sim (z)^2 / ((z')^2-z^2)^2$. In the presence of
spin flip scattering, one has $z \ne z'$ and
therefore a different conductance in the spin
flip and non spin flip channels.

\section{Conclusions}

To conclude, we have determined to what extend 
spin accumulation can result in an inverse Drude
behavior in a semi classical spin valve model.
Our treatment was based on an exact decoupling
between the charge and spin sectors of the
Boltzmann equation.
We have addressed a similar question in a
single impurity quantum model and found the
existence of a resonance in the spin flip
conductance. It is an open question
to determine the quantum coherent behavior
of a spin valve with a finite concentration
of impurities.

\appendix

\section{Derivation of the Boltzmann equation
with spin-flip scattering}
\begin{figure}
\centerline{\psfig{file=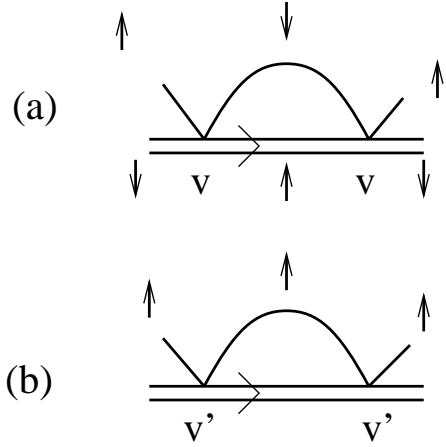,height=6cm}}
\caption{ The self energy terms incorporated in the
gradient expansion calculation. The term shown on
Fig.~\ref{fig:diagram1}~(b) generating the Kondo effect
is not included.
}
\label{fig:diagram2}
\end{figure}
We give a brief derivation of the Boltzmann equation
Eq.~\ref{eq:Boltzmann} in the presence
of a spin-flip scattering potential. 
The derivation
generalizes Ref.~\cite{Rammer86} to incorporate a
spin-flip scattering self energy. The Dyson equation
in the spin tensor Keldysh space reads
$(\hat{G}_0^{-1} - \hat{\Sigma})(1,2) \otimes \hat{G} (2)
= \delta(1-2)$. The convolution includes a sum over
coordinates, time, and spin. The kinetic equation
is obtained from the difference of the Keldysh
components of the Dyson equations and its conjugate:
\begin{equation}
\label{eq:kinetic}
 \left[ \hat{G}_0^{-1} - \mbox{Re} \hat{\Sigma} ,
\hat{G}^K \right]_-
- \left[ \hat{\Sigma}^K , \mbox{Re} \hat{G} \right]_-
= \frac{i}{2} \left[ \hat{\Sigma}^K , \hat{A} \right]_+
 - \frac{i}{2} \left[ \hat{\Gamma} , \hat{G}^K \right]_+
,
\end{equation}
with $\left[ \right]_-$ and $\left[ \right]_+$
denoting a commutator and an anticommutator
respectively. We refer the
reader to Ref.~\cite{Rammer86} for an explanation of
the symbols used in Eq.~\ref{eq:kinetic}.
We use the
the self energy shown on Fig.~\ref{fig:diagram2}:
\begin{equation}
\label{eq:self}
\hat{\Sigma}_{\sigma}({\bf p},{\bf R},T)
= n_i \int \frac{ d{\bf p}'}{(2 \pi)^3}
| v({\bf p} - {\bf p}')|^2 \hat{G}_{\sigma,\sigma}({\bf p}',
{\bf R},T)
+ n_i' \int \frac{ d {\bf p}'}{(2 \pi)^3}
|v'({\bf p} - {\bf p}')|^2 \hat{G}_{- \sigma,-\sigma}
({\bf p}', {\bf R}, T)
,
\end{equation}
with $n_i$ and $n_i'$ the concentration of non
magnetic and magnetic impurities.
The first term in Eq.~\ref{eq:self} describes non
spin-flip scattering, and the second term describes
spin-flip scattering. Notice that this self energy does
not incorporate the Kondo effect since we do not
incorporate the possibility of a having hole in the
intermediate state.

We assume the self energy
in Eq.~\ref{eq:kinetic} to be constant in space,
and use the gradient expansion to first order
$$
\left( A \otimes B \right)_{\sigma,\sigma'} (X,p)
\simeq \sum_{\sigma_2}
\left[ 1 + \frac{i}{2} \left( \partial_X^A
\partial_p^B - \partial_p^A \partial_X^B \right)
\right] A_{\sigma,\sigma_2} B_{\sigma_2,\sigma'}
.
$$
If $A$ and $B$ are symmetric in spin $A_{\sigma,\sigma'}=
A_{\sigma',\sigma}$, and $B_{\sigma,\sigma'}=
B_{\sigma',\sigma}$, the commutator reduces
to the usual spinless Poisson bracket: 
$\left[ A \otimes B \right]_{-,\sigma,\sigma'}
= i \sum_{\sigma_2}
\left\{A_{\sigma,\sigma_2} , B_{\sigma2,\sigma} \right\}$,
with $\left\{ A,B \right\} = \partial_X^A A \partial_p^B B
- \partial_p^A A \partial_X^B B$. Using these relations, we
expand the kinetic
equation Eq.~\ref{eq:kinetic}
and integrate over energy to obtain the Boltzmann equation
\begin{eqnarray}
\label{eq:B-spin-flip}
\partial_T f_{ {\bf p}, \sigma}
+ {\bf \nabla}_{\bf p} \xi_{\bf p} {\bf \nabla}_{\bf R}
f_{ {\bf p},\sigma} - {\bf \nabla}_{\bf R} U
{\bf \nabla}_{\bf p} f_{ {\bf p},\sigma}
&=& 2 \pi n_i \int \frac{ d {\bf p}'}{ (2 \pi)^3}
|v({\bf p} - {\bf p}')|^2 \delta( \xi_{\bf p}
- \xi_{ {\bf p}'} ) [f_{ {\bf p}', \sigma}
- f _{ {\bf p},\sigma} ] \\
&& + 2 \pi n_i' \int \frac{d {\bf p}'}{(2 \pi)^3}
| v'({\bf p} - {\bf p}')|^2 \delta ( \xi_{\bf p}
- \xi_{ {\bf p}'}) [ f_{ {\bf p}',-\sigma}
- f_{ {\bf p},\sigma}]
\nonumber
.
\end{eqnarray}
In the one dimensional limit,
Eq.~\ref{eq:B-spin-flip} reduces to the Boltzmann
equation Eq.~\ref{eq:Boltzmann}, with the scattering
coefficients related to the $q=0$ and $q = 2 k_f$
components of the scattering potential:
$r=n_i |v_{2 k_f}|^2$,
$r_s = n_i' |v_0|^2$, and $r_s'=n_i' |v_{2 k_f}'|^2$.


\newpage

\begin{thebibliography}{99}

\bibitem{exp1} M.N. Baibich, J.M. Broto, A. Fert,
F. Nguyen Van Dau, F. Petroff, P. Etienne,
G. Creuzet, A. Friederich, and J. Chazelas,
Phys. Rev. Lett. {\bf 61}, 2472 (1988).

\bibitem{exp2} G. Binach, P. Grunberg, F. Saurenbach,
and W. Zinn, Phys. Rev. B {\bf 39}, 4828 (1989).

\bibitem{exp3} W.P. Pratt,, Jr., S.F. Lee,
J.M. Slaughter, R. Loloee, P.A.
Schroeder, and J. Bass,
Phys. Rev. Lett. {\bf 66}, 3060 (1991);
S.F. Lee, W.P. Pratt, Jr., R. Loloee, P.A.
Schroeder, and J. Bass, Phys. Rev. B
{\bf 46}, 548 (1992).

\bibitem{Valet} T. Valet and A. Fert, Phys.
Rev. B {\bf 48}, 7099 (1993).

T. Valet and A. Fert, J. Magn. Magn.
Mater. {\bf 121}, 378 (1993);

A. Fert, T. Valet, and J. Barnas, J. Appl. Phys.
{\bf 75}, 6693 (1994).

A. Fert, J.L. Duvail, and T. Valet,
Phys. Rev. B {\bf 52}, 6513 (1995).

\bibitem{Bauer} M.A.M. Gijs and G. Bauer,
Adv. in Physics {\bf 46}, 285 (1997).

\bibitem{Falko} V.J. Falko, C.J. Lambert, and
A.F. Volkov, JETP Letters {\bf 69},
532 (1999).

\bibitem{Jedema} F.J. Jedema, B.J. van Wees,
B.H. Hoving, A.T. Filip, and T.M. Klapwijk,
Phys. Rev. B {\bf 60}, 16549 (1999).

\bibitem{Belzig} W. Belzig, A. Brataas, Yu. V. Nazarov,
G.E.W. Bauer,
arXiv:cond-mat/0005188.

\bibitem{Gu98} R.Y. Gu, D.Y. Xing, and Z.D. Wang,
Phys. Rev. B {\bf 58}, 11142 (1998).

\bibitem{Chen98} J. Chen and S. Hershfield,
Phys. Rev. B {\bf 57}, 1097 (1998).

\bibitem{Johnson} M. Johnson and R.H. Silsbee,
Phys. Rev. B {\bf 35}, 4959 (1987).

\bibitem{vanSon} P.C. Van Son, H. Van Kempen,
and P. Wyder, Phys. Rev. Lett. {\bf 58},
2271 (1987).

\bibitem{Zhu97} J.X. Zhu and Z.D. Wang,
Phys. Rev. B {\bf 55}, 8437 (1997).

\bibitem{Hewson93} {\sl The Kondo Problem to Heavy Fermions},
A.C. Hewson, Cambridge Studies in Magnetism, Cambridge
University Press (1993).


\bibitem{Kozub95} V.I. Kozub and A.M. Rudin, Phys. Rev. B
{\bf 52}, 7853 (1995).

\bibitem{Nagaev95} K.E. Nagaev, Phys. Rev. B {\bf 52},
4740 (1995).

\bibitem{Blonder82} G.E. Blonder, M. Tinkham, and T.M.
Klapwijk, Phys. Rev. B {\bf 25}, 4515 (1982).


\bibitem{Rammer86} J. Rammer and H. Smith, Rev. Mod.
Physics {\bf 58}, 323 (1986).


\end{thebibliography}
\end{document}